\documentclass[11pt]{article}
\usepackage{amsmath,amssymb,latexsym,graphicx}

\begin{document}

\begin{flushright}
	OUHEP-514 \\
	YITP-04-75 \\
	OIQP-04-05 \\
	hep-th/0501095
\end{flushright}

\vspace{0.5cm}

\begin{center}
{\Large\bf Negative Energy (Dirac-like) Sea for Bosons and Supersymmetry}\footnotemark \\

\footnotetext{To appear in Proceedings of the 7th Workshop``What comes
beyond standard model'', Bled, Slovenia, 19-30 July, 2004}

\vspace{0.5cm}
{\large Yoshinobu H}ABARA\\
\vspace{0.2cm}
{\it  Department of Physics, Graduate School of Science,}\\
{\it  Osaka University, Toyonaka, Osaka 560-0043, Japan}\\
\vspace{1cm}
{\large Holger B. N}IELSEN\\
\vspace{0.2cm}
{\it  Niels Bohr Institute, University Copenhagen,}\\
{\it  17 Blegdamsvej Copenhagen \o, Denmark}\\
\vspace{0.3cm}
and\\
\vspace{0.3cm}
{\large Masao N}INOMIYA\footnotemark \\
\vspace{0.2cm}
{\it  Yukawa Institute for Theoretical Physics,}\\
{\it  Kyoto University, Kyoto 606-8502, Japan}
\end{center}

\footnotetext{After April 1, 2004, also working at Okayama Institute for Quantum Physics, Kyoyama-cho 1-9, Okayama-city 700-0015, Japan}

\vspace{0.5cm}

\begin{abstract}
We consider the long standing problem in field theories of bosons that the boson vacuum does not consist of a `sea', unlike the fermion vacuum. We show with the help of supersymmetry considerations that the boson vacuum indeed does also consist of a sea in which the negative energy states are all ``filled", analogous to the Dirac sea of the fermion vacuum, and that a hole produced by the annihilation of one negative energy boson is an anti-particle. This might be formally coped with by introducing the notion of a double harmonic oscillator, which is obtained by extending the condition imposed on the wave function. Next, we present an attempt to formulate the supersymmetric and relativistic quantum mechanics utilizing the equations of motion.
\end{abstract}

\newpage

%%%%%%%%%%%%%%%%%%%%%%%%%%%%%%%%%%%%%%%%%%%%%%%%%%%%%%%%%%%%%%%%%%%%%%%%%%%%%%%
\section{Introduction}

In the Quantum Field Theory there is a long-standing historical mystery why there is no negative energy sea for bosons, contrary to fermions when we pass from 1st quantized theory into 2nd quantized one. Needless to say nowadays one uses a well functioning method that in the negative energy states creation operator and destruction operators should be formally exchanged. This rewriting can be used both bosons and fermions. In this formal procedure, as for fermions, the true vacuum is the one that the negative energy states are completely filled for one particle in each state due to Pauli principle. By filling all empty negative states to form Dirac sea \cite{dirac} we define creation operators $\tilde{d}^+(\vec{p},s,w)$ with positive energy $w$ for holes which is equivalent to destruction operators $d(-\vec{p},-s,-w)$. In particular for boson case the associated filling of the negative energy state to form a sea as the true vacuum has never been heard.\footnote{The interesting historical description can be found in ref.\cite{weinberg}}

In this report one of the main new contents present a new method of 2nd quantization of boson field theories in the analogous manner to filling of empty Dirac sea for fermions.

At the very end when the true vacuum in the 2nd quantized boson theory is formed according to our method\cite{nn,hnn} we will come exactly the same theory as the usual one. Thus our approach cannot be incorrect as far of quantum field theories are concerned. However application to the string theories may be very interesting open question. Although we have to introduce such an unfamiliar notion to form negative energy sea (Dirac) sea for boson that we have to subtract ``minus one boson'' or create ``a boson hole" in boson sea.

Validity and importance of our new method to 2nd quantize boson theories may become more clear by considering supersymmetric field theories. Due to supersymmetry one should expect the boson vacuum structure similar to the fermion one. We in fact utilize the N=2 matter multiplet called a hypermultiplet\cite{sohnius} and construct the Noether current from the supersymmetric action. By requiring that the entire system be supersymmetric, we derive the properties of the boson vacuum, while the fermion vacuum is taken to be the Dirac sea.

In the supersymmetric theories there seems to exist no truly quantum mechanical relativistic, i.e. 1 particle model so that one starts from the field theories. We in the present article propose a method to construct sypersymmetric quantum mechanics, although we have not yet completed the procedure~\cite{hnn2}.

The present article is organized as follows: In the next section 2 we utilize the N=2 supersymmetry to further clarify the vacuum structures of boson as well as fermions; both vacuua should be expected to be very similar by the supersymmetry. We utilize N=2 matter multiplet called hypermultiplet and explicitly show that the boson true vacuum can be formulated so as to form the negative energy sea, i.e. Dirac sea. In section 3 we start with describing the harmonic oscillator with the requirement of analyticity of the wave function instead of the usual square integrability condition. This naturally leads to the boson negative energy (Dirac sea) states as well as the usual positive ones. We then argue how to treat the Dirac sea for bosons. In the next section 4 we turn our attention to the 1 particle supersymmetric theory which may be considered 1st excited states in the boson vacuum, i.e. filled Dirac seas for bosons as well as fermions in our formalism. This theory is viewed as supersymmetric relativistic quantum mechanics. We propose a prescription how to realize the supersymmetry in a single particle level. The final section 5 is devoted to the conclusions.

%%%%%%%%%%%%%%%%%%%%%%%%%%%%%%%%%%%%%%%%%%%%%%%%%%%%%%%%%%%%%%%%%%%%%%%%%%%%%%%
\section{Boson and Dirac Seas in a Hypermultiplet Model with $N=2$ Supersymmetry}

In the present section, we consider an ideal world in which supersymmetry holds exactly. Then, it is natural to believe that, in analogy to the true fermion vacuum, the true boson vacuum is a state in which all negative energy states are occupied. To investigate the details of the vacuum structure of bosons, we utilize the $N=2$ matter multiplet called a hypermultiplet~\cite{sohnius,west}. In fact, we construct the Noether current from the supersymmetric action, and by requiring that the entire system be supersymmetric, we derive the properties of the boson vacuum, while the fermion vacuum is taken to be the Dirac sea.

Hereafter, the Greek indices $\mu ,\nu ,\cdots$ are understood to run from 0 to 3, corresponding to the Minkowski space, and the metric is given by $\eta^{\mu \nu}=diag(+1,-1,-1,-1)$.

%%%%%%%%%%%%%%%%%%%%%%%%%%%%%%%%%%%%%%%%%%%%%%%%%%%%%%%%%%%%%%%%%%%%%%%%%%%%%%%
\subsection{$N=2$ matter multiplet: Hypermultiplet}

Let us summarize the necessary part of $N=2$ supersymmetric field theory in the free case, in order to reveal the difficulty of making supersymmetric description. The hypermultiplet is the simplest multiplet that is supersymmetric and involves a Dirac fermion. It is written 

\begin{align*}
	\phi =(A_1, A_2;\> \psi ;\> F_1, F_2), \tag{2.1}
\end{align*}

\noindent where $A_i$ and $F_i\> (i=1,2)$ denote complex scalar fields, and the Dirac field is given by $\psi$. The multiplet (2.1) transforms under a supersymmetric transformation as 

\begin{align*}
	& \delta_{\xi}A_i=2\bar{\xi}_i\psi , \\
	& \delta_{\xi}\psi =-i\xi_iF_i-i\gamma^{\mu}\partial_{\mu}\xi_iA_i, \\
	& \delta_{\xi}F_i=2\bar{\xi_i}\gamma^{\mu}\partial_{\mu}\psi ,\tag{2.2}
\end{align*}

\noindent where $\gamma^{\mu}$ denotes the four-dimensional gamma matrices, with $\{ \gamma^{\mu},\gamma^{\nu}\} =2\eta^{\mu \nu}$. Then, we obtain the Lagrangian density of the hypermultiplet, 

\begin{align}
	\mathcal{L} & =\frac{1}{2}\partial_{\mu}A_i^{\dagger}\partial^{\mu}
	A_i+\frac{1}{2}F_i^{\dagger}F_i+\frac{i}{2}\bar{\psi}\gamma^{\mu}
	\partial_{\mu}\psi -\frac{i}{2}\partial_{\mu}\bar{\psi}\gamma^{\mu}
	\psi +m\left[\frac{i}{2}A_i^{\dagger}
	F_i-\frac{i}{2}F_i^{\dagger}A_i+\bar{\psi}\psi \right]. \tag{2.4}
\end{align}

To derive the Noether currents whose charges generate the supersymmetry transformation, we consider a variation under the supersymmetry transformation (2.2), 

\begin{align*}
	\delta_{\xi}\mathcal{L}=\bar{\xi}_i\partial_{\mu}K_i^{\mu}
	+\partial_{\mu}\bar{K}_i^{\mu}\xi_i, \tag{2.5}
\end{align*}

\noindent where $K_i^{\mu}$ is given by 

\begin{align*}
	K_i^{\mu}\equiv \frac{1}{2}(\gamma^{\mu}\gamma^{\nu}\psi 
	\partial_{\nu}A_i^{\dagger}+im\gamma^{\mu}\psi A_i^{\dagger}). 
	\tag{2.6}
\end{align*}

\noindent Thus, the Noether current $J_i^{\mu}$ is written as 

\begin{align*}
	& \bar{\xi}_iJ_i^{\mu}+\bar{J}_i^{\mu}\xi_i=
	\frac{\delta L}{\delta_{\xi}(\partial_{\mu}\phi)}\delta_{\xi}\phi 
	-\left(\bar{\xi}_iK_i^{\mu}+\bar{K}_i^{\mu}\xi_i\right) \\
	& \quad \Longrightarrow J_i^{\mu}=\gamma^{\nu}\gamma^{\mu}\psi 
	\partial_{\nu}A_i^{\dagger}-im\gamma^{\mu}\psi A_i^{\dagger}. \tag{2.7}
\end{align*}

We could attempt to think of treating the bosons analogous to the fermions by imagining that the creation operators $a^{\dagger}(\vec{k})$ of the anti-bosons were really annihilation operators in some other formulation, but, as we shall see, such an attempt leads to some difficulties. If we could indeed do so, we would write also a boson field in terms of only annihilation operators formally as follows 

\begin{align*}
	& A_i(x)=\int \frac{d^3\vec{k}}{\sqrt{(2\pi )^32k_0}}\left\{
	a_{i+}(\vec{k})e^{-ikx}+a_{i-}(\vec{k})e^{ikx}\right\}, \tag{2.8} \\
	& \psi(x)=\int \frac{d^3\vec{k}}{\sqrt{(2\pi )^32k_0}}\sum_{s=\pm}
	\left\{b(\vec{k},s)u(\vec{k},s)e^{-ikx}+d(\vec{k},s)v(\vec{k},s)e^{ikx}
	\right\}. \tag{2.9}
\end{align*}

\noindent Here, $k_0\equiv \sqrt{\vec{k}^2+m^2}$ is the energy of the particle, and $s\equiv \frac{\vec{\sigma}\cdot \vec{k}}{|\vec{k}|}$ denotes the helicity. Particles with positive and negative energy are described by $a_{i+}(\vec{k}), b(\vec{k},s)$ and $a_{i-}(\vec{k}), d(\vec{k},s)$, respectively. The commutation relations between these field modes are derived as 

\begin{align*}
	& \left[a_{i+}(\vec{k}),a_{j+}^{\dagger}(\vec{k}^{\prime})\right]=
	+\delta_{ij}\delta^3(\vec{k}-\vec{k}^{\prime}), 
	\quad \left[a_{i-}(\vec{k}),a_{j-}^{\dagger}(\vec{k}^{\prime})\right]=
	-\delta_{ij}\delta^3(\vec{k}-\vec{k}^{\prime}), \tag{2.10} \\
	& \left\{b(\vec{k},s),b^{\dagger}(\vec{k}^{\prime},s^{\prime})\right\}
	=+\delta_{ss^{\prime}}\delta^3(\vec{k}-\vec{k}^{\prime}), 
	\quad \left\{d(\vec{k},s),d^{\dagger}(\vec{k}^{\prime},s^{\prime})
	\right\}=+\delta_{ss^{\prime}}\delta^3(\vec{k}-\vec{k}^{\prime}), 
	\tag{2.11}
\end{align*}

\noindent with all other pairs commuting or anti-commuting. Note that the right-hand side of the commutation relation (2.10) for negative energy bosons, has the opposite sign of that for positive energy bosons. In the ordinary method, recalling that the Dirac sea is the true fermion vacuum, we can use $d^{\dagger}$ as the creation operator and $d$ as the annihilation operator for negative energy fermions. Then, the operators $d^{\dagger}$ and $d$ are re-interpreted as the annihilation operator and creation operator for positive energy holes. In this manner, we obtain the particle picture in the real world. In this procedure, negative energy fermions are regarded as actually existing entities. 

For bosons, in contrast to the fermions, we rewrite the second equation of (2.10) as 

\begin{align*}
	& \> a_{i-}\equiv \tilde{a}_{i-}^{\dagger}, \quad a_{i-}^{\dagger}
	\equiv \tilde{a}_{i-}, \\
	& \left[\tilde{a}_{i-}(\vec{k}),\tilde{a}_{j-}^{\dagger}
	(\vec{k}^{\prime})\right]=+\delta_{ij}\delta^3
	(\vec{k}-\vec{k}^{\prime}). \tag{2.12}
\end{align*}

\noindent This implies that we can treat negative energy bosons in the same manner as positive energy bosons. Consequently, the true vacua for positive and negative energy bosons, which are denoted $||0_+\rangle$ and $||0_-\rangle$, respectively\footnotemark, are given by 

\footnotetext{In the following, we denote the vacua by, for example in the boson case, $|0_{\pm}\rangle$ in the system of single particle, and $||0_{\pm}\rangle$ in the system with many particles.}

\begin{align*}
	& a_{i+}||0_+\rangle =0, \\
	& \tilde{a}_{i-}||0_-\rangle =0. \tag{2.13}
\end{align*}

\noindent Thus, in the true vacuum, meaning the one on which our experimental world is built, both the negative and positive energy vacua are empty when using the particle $a_{i+}$ and anti-particle $\tilde{a}_{i-}$ annihilation operators respectively. However, in order to have a supersymmetry relation to the analogous negative energy states for the fermions, we would have liked to consider, instead of $||0_-\rangle$, a vacuum so that it were empty with respect to the negative energy bosons described by $a_{i-}$ and $a_{i-}^{\dagger}$. That is to say we would have liked a empty vacuum obeying $a_{i-}||0_{\text{wanted}}\rangle =0$. Because of (2.13) it is, however, immediately seen that this $||0_{\text{wanted}}\rangle$ cannot exist. In true nature, we should rather be in a situation or a ``sector" in which we have a state with $a_{i-}^{\dagger}||0_-\rangle =\tilde{a}_{i-}||0_-\rangle =0$. It could be called a ``sector with a top" $||0_-\rangle$.

Perhaps the nicest way of describing this extension is by means of the double harmonic oscillator to be presented in Section 3 below, but let us stress that all we need is a formal extrapolation to also include the possibility of negative numbers of bosons.

%%%%%%%%%%%%%%%%%%%%%%%%%%%%%%%%%%%%%%%%%%%%%%%%%%%%%%%%%%%%%%%%%%%%%%%%%%%%%%%
\subsection{Supersymmetry invariant vacuum}

As described in Subsection 2.1, when considered in terms of supersymmetry, there is a difference between the boson and fermion pictures. In the present subsection, we give preliminary considerations to the problem determining the nature of a boson sea that would correspond to the Dirac sea for the fermion case. To this end, we impose the natural condition within the supersymmetric theory that the vacuum be supersymmetry invariant. 

We first rewrite the supersymmetry charges $\mathcal{Q}_i$ derived from the supersymmetry currents described by Eq.(2.7) in terms of the creation and annihilation operators as 

\begin{align*}
	\mathcal{Q}_i & =\int d^3\vec{x} J_i^0(x)=i\int d^3\vec{k} 
	\sum_{s=\pm} \left\{b(\vec{k},s)u(\vec{k},s)
	a_{i+}^{\dagger}(\vec{k})-d(\vec{k},s)v(\vec{k},s)a_{i-}^{\dagger}
	(\vec{k})\right\}, \\
	\bar{\mathcal{Q}}_i & =\int d^3\vec{x} \bar{J}_i^0(x)=-i\int 
	d^3\vec{k} \sum_{s=\pm} \left\{b^{\dagger}(\vec{k},s)
	\bar{u}(\vec{k},s)a_{i+}(\vec{k})-d^{\dagger}(\vec{k},s)\bar{v}
	(\vec{k},s)a_{i-}(\vec{k})\right\}. \tag{2.14}
\end{align*}

\noindent By applying these charges, the condition for the vacuum to be supersymmetric can be written 

\begin{align*}
	\mathcal{Q}_i||0\rangle=\bar{\mathcal{Q}}_i||0\rangle=0. \tag{2.15}
\end{align*}

\noindent We then decompose the total vacuum into the boson and fermion vacua, $||0_{\pm}\rangle$ and $||\tilde{0}_{\pm}\rangle$, writing 

\begin{align*}
	||0\rangle \equiv ||0_+\rangle \otimes ||0_-\rangle \otimes 
	||\tilde{0}_+\rangle \otimes ||\tilde{0}_-\rangle , \tag{2.16}
\end{align*}

\noindent where $\otimes$ denotes the direct product, and $||\tilde{0}_-\rangle$ is the Dirac sea, given by 

\begin{align*}
	||\tilde{0}_-\rangle =\bigg\{\prod_{\vec{p},s}d^{\dagger}
	(\vec{p},s)\bigg\}||\tilde{0}\rangle .
\end{align*}

\noindent Here, $||\tilde{0}_+\rangle$ represents an empty vacuum, annihilated by the ordinary $b$ operator, while $||\tilde{0}_-\rangle$, given by Eq.(23), represents the Dirac sea, which is obtained through application of all $d^{\dagger}$. The condition for the bosonic vacuum reads 

\begin{align*}
	& a_{i+}(\vec{k})||0_+\rangle =0,\quad a_{i-}^{\dagger}(\vec{k})
	||0_-\rangle =0. \tag{2.17}
\end{align*}

\noindent It is evident that the vacuum of the positive energy boson $||0_+\rangle$ is the empty one, vanishing under the annihilation operator $a_{i+}$. On the other hand, the vacuum of the negative energy boson $||0_-\rangle$ is defined such that it vanishes under the operator $a_{i-}^{\dagger}$ that creates the negative energy quantum. This may seem very strange. One could call the strange ``algebra" looked for a ``sector with top", contrary to the more usual creation and annihilation systems which could rather be called ``sectors with a bottom".

In the next section, using the fact that the algebras (2.10) constitute that is essentially a harmonic oscillator system with infinitely many degrees of freedom, we investigate in detail the vacuum structure by considering the simplest one-dimensional harmonic oscillator system. In fact, we will find the explicit form of the vacuum $||0_-\rangle$ that is given by a coherent state of the excited states of all the negative energy bosons.

%%%%%%%%%%%%%%%%%%%%%%%%%%%%%%%%%%%%%%%%%%%%%%%%%%%%%%%%%%%%%%%%%%%%%%%%%%%%%%%
\section{Negative Energy (Dirac-like) Sea for Bosons}

When looking for solutions to the Klein-Gordon equation for energy (and momentum) it is well-known that, we must consider not only the positive energy particles but also the negative energy ones. In the previous section, we found that in order to implement the analogy to the Dirac sea for fermions suggested by supersymmetry, we would have liked to have at our disposal the possibility to organize an analogon of the Dirac sea (for fermions). In the present section we introduce the concept of a ``sector with top" as an extension of the harmonic oscillator spectrum to a negative energy sector. Thereby we have to extend the ordinary meaning of the wave function (in this case for the harmonic oscillator). Performing this we find that the vacuum of the negative energy sector leads to a ``boson sea", corresponding to the Dirac sea of fermions.

%%%%%%%%%%%%%%%%%%%%%%%%%%%%%%%%%%%%%%%%%%%%%%%%%%%%%%%%%%%%%%%%%%%%%%%%%%%%%%%
\subsection{Analytic wave function and double harmonic oscillator}

As is well known, the eigenfunction $\phi (x)$ of a one-dimensional Schr\"{o}dinger equation in the usual treatment should satisfy the square integrability condition, 

\begin{align*}
	\int_{-\infty}^{+\infty}dx \> |\phi (x)|^2<+\infty . \tag{3.1}
\end{align*}

\noindent If we apply this condition to a one-dimensional harmonic oscillator, we obtain as the vacuum solution only the empty one satisfying 

\begin{align*}
	a_+|0\rangle =0.
\end{align*}

\noindent Thus, we are forced to extend the condition for physically allowed wave functions in order to obtain ``boson sea" analogous to the Dirac sea. In fact we extend the condition (3.1), replacing it by the condition under which, when we analytically continuate $x$ to the entire complex plane, the wave function $\phi (x)$ is analytic and only an essential singularity is allowed as $|x|\! \to \! \infty$. In fact, for the harmonic oscillator, we can prove the following theorem: 

\vspace{0.5cm}

\noindent i) The eigenvalue spectrum $E$ for the harmonic oscillator 

\begin{align*}
	\left( -\frac{1}{2}\frac{d^2}{dx^2}+\frac{1}{2}x^2 \right) 
	\phi (x)=E\phi (x). \tag{3.2}
\end{align*}

\noindent is given by 

\begin{align*}
	E=\pm \left(n+\frac{1}{2}\right), \quad n\in \mathcal{Z}_+\cup \{0\}. 
	\tag{3.3}
\end{align*}

\noindent ii) The wave functions for positive energy states are the usual ones 

\begin{align*}
	& \phi_n(x)=A_nH_n(x)e^{-\frac{1}{2}x^2}, \nonumber \\
	& E=n+\frac{1}{2},\qquad n=0_+,1,2,\cdots . \tag{3.4}
\end{align*}

\noindent Here $H_n(x)$ is the Hermite polynomial while $A_n=\left( \sqrt{\pi}2^nn!\right)^{-\frac{1}{2}}$. For negative energy states, the eigenfunctions are given by 

\begin{align*}
	& \phi_{-n}(x)=A_nH_n(ix)e^{+\frac{1}{2}x^2}, \\
	& E=-\left( n+\frac{1}{2}\right) ,\qquad n=0_-,1,2,\cdots . \tag{3.5}
\end{align*}

\noindent iii) The inner product is defined as 

\begin{align*}
	\langle n|m\rangle =\int_{\Gamma}dx \> \phi_n(x^{\ast})^{\ast}
	\phi_m(x), \tag{3.6}
\end{align*}

\noindent where the contour is denoted by $\Gamma$. The $\Gamma$ should be chosen so that the integrand should go down to zero at $x=\infty$, but there remains some ambiguity in the choice of $\Gamma$. However if one chooses the same $\Gamma$ for all negative $n$ states, the norm of these states have an alternating sign.

\vspace{0.5cm}

The above i)-iii) constitute the theorem. Proof of this theorem is rather trivial, and we skip it by referring the refs.~\cite{nn,hnn}, but some comments are given in the following.

Going from (3.4) to (3.5) corresponds to the replacement $x\! \to \! ix$, so the creation and annihilation operators are transformed as 

\begin{align*}
	[a_+,a_+^{\dagger}]=+1 & \to [a_-,a_-^{\dagger}]=-1.
\end{align*}

\noindent Here, $a_+$ and $a_+^{\dagger}$ are ordinary operators in the positive energy sector, and $a_-$ and $a_-^{\dagger}$ are operators in the negative energy sector which create and annihilate negative energy quanta respectively.

It is useful to summarize the various results obtained to this point in operator form. We write each vacuum and excited state in the positive and negative energy sectors, respectively, as 

\begin{align*}
	& \phi_{+0}(x)=e^{-\frac{1}{2}x^2}\simeq |0_+\rangle , \tag{3.7} \\
	& \phi_{-0}(x)=e^{+\frac{1}{2}x^2}\simeq |0_-\rangle , \tag{3.8} \\
	& \phi_n(x)\simeq |n\rangle , \qquad n\in \mathcal{Z}-\{0\}. \tag{3.9}
\end{align*}

The important point here is that there exists a gap between the positive and negative sectors. Suppose that we write the states in order of their energies as 

\begin{figure}[ht]
	\begin{center}
	\begin{picture}(360,40)
	\put(40,20){$\cdots$}
	\put(65,23){$\xrightarrow[]{a^{\dagger}}$}
	\put(65,14){$\xleftarrow[\> a \>]{}$}
	\put(85,18){$|\! -\! 1\rangle$}
	\put(115,23){$\xrightarrow[]{a^{\dagger}}$}
	\put(115,14){$\xleftarrow[\> a \>]{}$}
	\put(135,18){$|0_-\rangle$}
	\put(160,23){$\xrightarrow[]{a^{\dagger}}$}
	\put(180,18){$0$}
	\put(190,14){$\xleftarrow[\> a \>]{}$}
	\put(210,18){$|0_+\rangle$}
	\put(235,23){$\xrightarrow[]{a^{\dagger}}$}
	\put(235,14){$\xleftarrow[\> a \>]{}$}
	\put(255,18){$|\! +\! 1\rangle$}
	\put(285,23){$\xrightarrow[]{a^{\dagger}}$}
	\put(285,14){$\xleftarrow[\> a \>]{}$}
	\put(310,20){$\cdots$.}
	\end{picture}
	\end{center}
	\label{sequence}
\end{figure}%

\noindent As usual, the operators causing transitions in the right and left directions are $a_{\pm}^{\dagger}$ and $a_{\pm}$, respectively. However, between the two vacua $|0_-\rangle$ and $|0_+\rangle$ there is a ``wall" of the classical number $0$, and due to its presence, these two vacua cannot be transformed into each other under the operations of $a_{\pm}$ and $a_{\pm}^{\dagger}$. In going to the second quantized theory with interactions, there appears to be the possibility of such a transition. However, it turns out that the usual polynomial interactions do not induce such a transition.

Next, we comment on the definition of the inner product of states. As explained above, there exists a gap such that no transition between the two sectors can take place. Thus, we can define the inner product of only states in the same sector. The inner product that in the positive energy sector provides the normalization condition is, as usual, given by 

\begin{align}
	\langle n|m\rangle & \equiv \int_{-\infty}^{+\infty}dx\> 
	\phi_n^{\dagger}(x)\phi_m(x)=\delta_{nm}, \qquad n,m=0_+,1,2,\cdots . 
	\tag{3.10}
\end{align}

\noindent However, the eigenfunctions in the negative energy sector are obtained as Eq.(3.5), so we propose  a path of integration such that the integration is convergent, since we impose the condition that the wave functions are analytic. Then, we define the inner product in terms of a path $\Gamma$, which we make explicit subsequently: 

\begin{align}
	\langle n|m\rangle \equiv -i\int_{\Gamma}dx\> \phi_n^{\ast}(x^{\ast})
	\phi_m(x)=(-1)^n\delta_{nm}, \qquad n,m=0_-,-1,-2,\cdots \tag{3.11}
\end{align}

\begin{figure}[ht]
	\begin{center}
	\begin{picture}(180,180)
	\put(0,90){\vector(1,0){180}}
	\put(90,0){\vector(0,1){180}}
	\put(170,170){\line(1,0){10}}
	\put(170,170){\line(0,1){10}}
	\put(172,172){$x$}
	\put(75,81){$0$}
	\put(172,95){Re$\> x$}
	\put(62,172){Im$\> x$}
	\put(15,15){\line(1,1){150}}
	\put(15,165){\line(1,-1){150}}
	\put(80,100){\line(1,0){20}}
	\put(70,110){\line(1,0){40}}
	\put(60,120){\line(1,0){60}}
	\put(50,130){\line(1,0){80}}
	\put(40,140){\line(1,0){100}}
	\put(30,150){\line(1,0){120}}
	\put(20,160){\line(1,0){140}}
	\put(80,80){\line(1,0){20}}
	\put(70,70){\line(1,0){40}}
	\put(60,60){\line(1,0){60}}
	\put(50,50){\line(1,0){80}}
	\put(40,40){\line(1,0){100}}
	\put(30,30){\line(1,0){120}}
	\put(20,20){\line(1,0){140}}
	\qbezier(90,90)(50,130)(130,170)
	\qbezier(90,90)(130,50)(50,10)
	\put(120,170){$\Gamma$}
	\put(98,150){\vector(4,3){10}}
	\end{picture}
	\end{center}
	\caption[path $\Gamma$]{Path $\Gamma$ for which the inner product 
	(3.11) converges.}
	\label{gamma}
\end{figure}
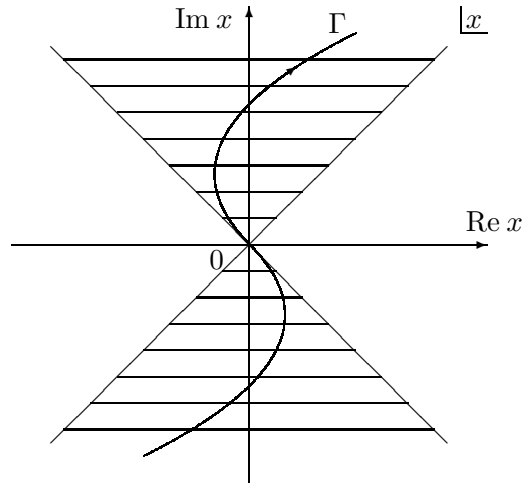%

\noindent Here, it is understood that the complex conjugation yielding $\phi^{\ast}(x^{\ast})$ is taken so that the inner product is invariant under deformations of the path $\Gamma$ within the same topological class in the lined regions as shown in Fig.~\ref{gamma}.

Therefore, the definition of the inner product in the negative energy sector is not essentially different from that of the positive energy sector, except for the result of the alternating signature $(-1)^{n}\delta_{nm}$, which is, however, very crucial. On the other hand, if we adopt $\phi^{\ast}(x)$ instead of $\phi^{\ast}(x^{\ast})$ in (3.11), we can also obtain the positive definite inner product $\langle n|m\rangle =\delta_{nm}$ for the negative energy sector at the sacrifice of the path-independence of $\Gamma$.

%%%%%%%%%%%%%%%%%%%%%%%%%%%%%%%%%%%%%%%%%%%%%%%%%%%%%%%%%%%%%%%%%%%%%%%%%%%%%%%
\subsection{Boson vacuum in the negative energy sector}

The vacua $|0_+\rangle$ and $|0_-\rangle$ in the positive and negative energy sectors are 

\begin{align*}
	& |0_+\rangle \simeq e^{-\frac{1}{2}x^2}, \tag{3.12} \\
	& |0_-\rangle \simeq e^{+\frac{1}{2}x^2}. \tag{3.13}
\end{align*}

\noindent In order to demonstrate how $|0_-\rangle$ represents a sea, we derive a relation between the two vacua (3.12) and (3.13) analogous to that in the fermion case. In fact, by comparing the explicit functional forms of each vacuum, we easily find the relation 

\begin{align*}
	& e^{+\frac{1}{2}x^2}=e^{x^2}\cdot e^{-\frac{1}{2}x^2}, \qquad 
	e^{x^2}=e^{\frac{1}{2}(a+a^{\dagger})^2} \nonumber \\
	& \Longrightarrow 
	|0_-\rangle =e^{\frac{1}{2}(a+a^{\dagger})^2}|0_+\rangle . \tag{3.14}
\end{align*}

\noindent This relation is preferable for bosons for the following reason. In the fermion case, due to the exclusion principle, the Dirac sea is obtained by exciting only one quantum of the empty vacuum. Contrastingly, because in the boson case there is no exclusion principle, the vacuum $|0_-\rangle$ in the negative energy sector is constructed as a sea by exciting all even number of quanta, i.e. an infinite number of quanta. 

%%%%%%%%%%%%%%%%%%%%%%%%%%%%%%%%%%%%%%%%%%%%%%%%%%%%%%%%%%%%%%%%%%%%%%%%%%%%%%%
\subsection{Boson sea}

In the present subsection, we investigate the boson vacuum structure in detail, utilizing the second quantized theory for a complex scalar field. Firstly, we clarify the properties of the unfamiliar vacuum $||0_-\rangle$ in the negative energy sector, using the result of Subsection 3.1. To this end, we study the details of the infinite-dimensional harmonic oscillator, which is identical to a system of a second quantized complex scalar field. The representation of the algebra (2.10) that is formed by $a_+,a_-$ and their conjugate operators is expressed as 

\begin{align*}
	& a_+(\vec{k})=\left( A(\vec{k})+\frac{\delta}{\delta A(\vec{k})}
	\right), \quad a_+^{\dagger}(\vec{k})=\left( A(\vec{k})
	-\frac{\delta}{\delta A(\vec{k})}\right), \tag{3.15} \\
	& a_-(\vec{k})=i\left( A(\vec{k})+\frac{\delta}{\delta A(\vec{k})}
	\right),\quad a_-^{\dagger}(\vec{k})=i\left( A(\vec{k})
	-\frac{\delta}{\delta A(\vec{k})}\right). \tag{3.16}
\end{align*}

\noindent The Hamiltonian and Schr\"{o}dinger equation of this system as the infinite-dimensional harmonic oscillator read 

\begin{align*}
	& H=\int \frac{d^3\vec{k}}{(2\pi )^3}\left\{ -\frac{1}{2}
	\frac{\delta^2}{\delta A^2(\vec{k})}+\frac{1}{2}A^2(\vec{k})\right\} 
	, \tag{3.17} \\
	& H\Phi [A]=E\Phi [A]. \tag{3.18}
\end{align*}

\noindent Here, $\Phi [A]$ denotes a wave functional of the wave function $A(\vec{k})$. We are now able to write an explicit wave functional for the vacua of the positive and negative enegy sectors: 

\begin{align*}
	& ||0_+\rangle \simeq \Phi_{0_+}[A]=e^{-\! \frac{1}{2} \int \! 
	\frac{d^3\vec{k}}{(2\pi )^3}A^2(\vec{k})}, \tag{3.19} \\
	& ||0_-\rangle \simeq \Phi_{0_-}[A]=e^{+\! \frac{1}{2} \int \! 
	\frac{d^3\vec{k}}{(2\pi )^3}A^2(\vec{k})}. \tag{3.20}
\end{align*}

\noindent We can find a relation between these two vacua via Eq.(3.14): 

\begin{align*}
	||0_-\rangle & =e^{\int \! \frac{d^3\vec{k}}{(2\pi )^3}A^2(\vec{k})}
	||0_+\rangle \nonumber \\
	& =e^{-\frac{1}{2}\! \int \! \frac{d^3\vec{k}}{(2\pi )^3}
	\left\{ a_-(\vec{k})+a_-^{\dagger}(\vec{k})\right\}^2}||0_+\rangle . 
	\tag{3.21}
\end{align*}

\noindent From this equation, we see that the negative energy vacuum $||0_-\rangle$ is a coherent state constructed from the empty vacuum $||0_+\rangle$ of the positive energy sector by creating all the even number negative energy bosons through the action of $a_-^{\dagger}(\vec{k})$. In this sense, $||0_-\rangle$ is the sea in which all the negative energy boson states are filled.

\begin{figure}[ht]
	\begin{center}
	\begin{picture}(200,160)
	\put(35,0){\line(0,1){140}}
	\put(50,0){\line(0,1){140}}
	\put(150,0){\line(0,1){140}}
	\put(165,0){\line(0,1){140}}
	\put(93,67){\Large $\otimes$}
	\put(40,-10){\Large $\vdots$}
	\put(40,140){\Large $\vdots$}
	\put(155,-10){\Large $\vdots$}
	\put(155,140){\Large $\vdots$}
	\put(-25,130){\scriptsize positive energy}
	\put(-5,120){\scriptsize sector}
	\put(185,20){\scriptsize negative energy}
	\put(205,10){\scriptsize sector}
	\multiput(35,75)(0,10){7}{\line(1,0){15}}
	\multiput(150,5)(0,10){7}{\line(1,0){15}}
	\multiput(0,70)(10,0){9}{\line(1,0){5}}
	\multiput(115,70)(10,0){9}{\line(1,0){5}}
	\put(53,73){\scriptsize $0_+$}
	\put(53,83){\scriptsize $+1$}
	\put(53,93){\scriptsize $+2$}
	\put(53,103){\scriptsize $+3$}
	\put(53,113){\scriptsize $+4$}
	\put(53,123){\scriptsize $+5$}
	\put(53,133){\scriptsize $+6$}
	\put(168,63){\scriptsize $0_-$}
	\put(168,53){\scriptsize $-1$}
	\put(168,43){\scriptsize $-2$}
	\put(168,33){\scriptsize $-3$}
	\put(168,23){\scriptsize $-4$}
	\put(168,13){\scriptsize $-5$}
	\put(168,3){\scriptsize $-6$}
	\put(205,67){wall of $0$ (zero)}
	\end{picture}
	\end{center}
	\caption[tower of states]{Physical states in the two sectors.}
	\label{tower}
\end{figure}
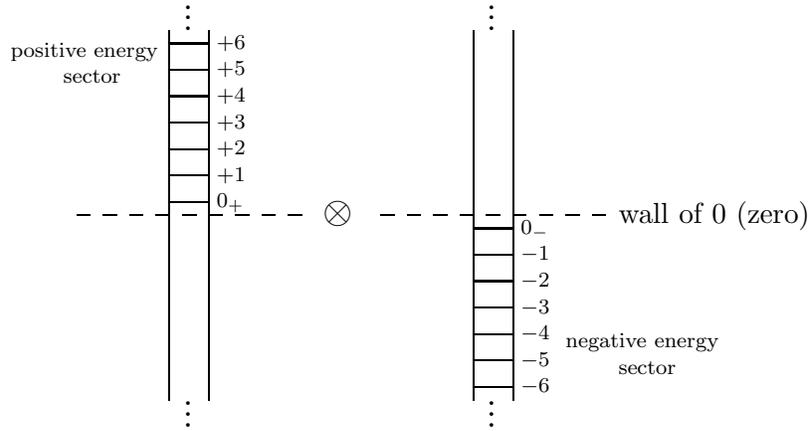

To avoid the misconceptions that the positive and negative energy sectors may simultaneously coexist and that there is no distinction between them, we depict them in Fig.~\ref{tower}.

To end the present section, a comment about the inner product of the states in the second quantized theory is in order. If we write the $n$-th excited state as $||n\rangle \simeq \Phi_n[A]$, the inner product is defined by 

\begin{align}
	\langle n||m\rangle =\int \mathfrak{D}A\> \Phi_n^{\ast}[A^{\ast}]
	\Phi_m[A], \tag{3.22}
\end{align}

\noindent where on the right-hand side there appears a functional integration over the scalar field $A(\vec{k})$. Recalling the definition of a convergent inner product in the first quantization, (3.11), it might be thought that the integration over $A$ should be properly taken for $n,m=0_-,-1,-2,\cdots$ in order to make the integration convergent.

%%%%%%%%%%%%%%%%%%%%%%%%%%%%%%%%%%%%%%%%%%%%%%%%%%%%%%%%%%%%%%%%%%%%%%%%%%%%%%%
\section{Towards Supersymmetric Relativistic Quantum \\Mechanics}

In the proceeding section 3 we investigated the structure of the vacuua in the multiparticle system, i.e. field theory in terms of the supersymmetry. The present section we attempt to formulate the realization of the supersymmetric 1 particle system, i.e. relativistic quantum mechanics. This formulation in our 2nd qunantization method is really starting point: the 1 particle system appears as the excitation of the unfilled Dirac seas for bosons and fermions.

We first express the supersymmetry transformation for the chiral multiplet in the N=2 supersymmetric theory, 

\begin{align*}
	& \delta_{\xi}A=\sqrt{2}\xi \psi , \\
	& \delta_{\xi}\psi =i\sqrt{2}\sigma^{\mu}\bar{\xi}\vec{\partial}_{\mu}A
	+\sqrt{2}\xi F, \tag{4.1}\\
	& \delta_{\xi}F=i\sqrt{2}\bar{\xi}\bar{\sigma}^{\mu}
	\vec{\partial}_{\mu}\psi ,
\end{align*}

\noindent in terms of the matrices. For this purpose we write the chiral multiplet as the vector notation: 

\begin{align*}
	\vec{\Phi}=
	\left( \begin{array}{c}
	A \\ \psi_{\beta} \\ F
	\end{array} \right). \tag{4.2}
\end{align*}

Then the supersymmetric generators at the 1st quantized level in an off-shell representation which induces the SUSY transformation (4.1) is given by 

\begin{align*}
	Q_{\alpha}=
	\left( \begin{array}{ccc}
	0 & \sqrt{2}\delta_{\alpha}^{\> \beta} & 0 \\
	0 & 0 & \sqrt{2}\epsilon_{\beta \alpha} \\
	0 & 0 & 0
	\end{array} \right), \quad 
	\bar{Q}_{\dot{\alpha}}=
	\left( \begin{array}{ccc}
	0 & 0 & 0 \\
	i\sqrt{2}\sigma^{\mu}_{\beta \dot{\alpha}}\vec{\partial}_{\mu} & 0 & 0 
	\\
	0 & i\sqrt{2}\bar{\sigma}^{\mu \dot{\gamma}\beta}
	\epsilon_{\dot{\gamma}\dot{\alpha}}\vec{\partial}_{\mu} & 0
	\end{array} \right). \tag{4.3}
\end{align*}

The Lorentz-invariant inner product is defined as natural one, 

\begin{align*}
	& \langle \Phi |\Phi \rangle \equiv \int d^3\vec{x} \> 
	\vec{\Phi}^{\dagger}I\vec{\Phi}, \tag{4.4}
\end{align*}

\noindent with a matrix I,

\begin{align*}
	& I\equiv \left( \begin{array}{ccc}
	i\overleftrightarrow{\partial}^0 & 0 & 0 \\
	0 & -\bar{\sigma}^{0\dot{\alpha}\alpha} & 0 \\
	0 & 0 & 0 \\
	\end{array} \right). \tag{4.5}
\end{align*}

It is easy to check that the SUSY algebra $\left\{Q_{\alpha},\bar{Q}_{\dot{\alpha}}\right\}=i\sigma^{\mu}_{\alpha\dot{\alpha}}\vec{\partial}_{\mu}$ as well as hermiticity condition $\left<Q\Phi|\Phi\right>_{\dot{\alpha}}=\left<\Phi|\bar{Q}\Phi\right>_{\dot{\alpha}}$ hold by means of the equations of motion 

\begin{align}
	\vec{\partial}^2 A=0, \quad \sigma^{\mu}\vec{\partial}_{\mu}\bar{\psi}
	=0, \quad F=0. \tag{4.6}
\end{align}

\noindent For instance, by utilizing the equations of motion 

\begin{align*}
	(Q_{\alpha})^{\dagger}I=I\bar{Q}_{\dot{\alpha}}, \tag{4.7}
\end{align*}

\noindent where the dagger $\dagger$ on the left hand side denotes the usual hermitian conjuate of the matrix.

In passing to the 2nd quantized theory, the SUSY generator $\mathcal{Q}$ is expressed as 

\begin{align}
	\xi^{\alpha}\mathcal{Q}_{\alpha}+\bar{\mathcal{Q}}_{\dot{\alpha}}
	\bar{\xi}^{\dot{\alpha}}=\int d^3\vec{x} \> \Phi^{\dagger}I
	(\xi^{\alpha}Q_{\alpha}+\bar{Q}_{\dot{\alpha}}\bar{\xi}^{\dot{\alpha}})
	\Phi \tag{4.8}
\end{align}

If we wish, we may use the Dirac representation by doubling the number of fields and through the equations of motion we rederive the $\mathcal{Q}_{\alpha}$ and $\bar{\mathcal{Q}}_{\dot{\alpha}}$ in terms of the creation and annihilation operators which coincide with the equations (2.14).

Under the action of this SUSY generator (4.8) the vacuum

\begin{align}
	||0_+\rangle =||0\rangle \otimes ||0_-\rangle 
	\otimes ||\tilde{0}_+\rangle \otimes ||\tilde{0}_-\rangle \tag{4.9}
\end{align}

\noindent is variant. It is worthwhile to notice that 1 particle states are related each other as 

\begin{align*}
	& ||+\! 1,\vec{k}\rangle =a_+^{\dagger}(\vec{k})||0_+\rangle 
	\Longleftrightarrow
	||+\! \tilde{1},\vec{k}\rangle =b^{\dagger}(\vec{k})||\tilde{0}_+
	\rangle , \\
	&||-\! 1,\vec{k}\rangle =a_-(\vec{k})||0_-\rangle \Longleftrightarrow
	||-\! \tilde{1},\vec{k}\rangle =d(\vec{k})||\tilde{0}_-\rangle . 
	\tag{4.10}
\end{align*}

\noindent Thus it is clear that the supersymmetry at the level of 1 particle states is perfectly realized.

The unsolved problem to the authors is to derive the classical action which should lead to the supersymmetric quantum mechanics~\cite{hnn2}. It is still unknown how to obtain this classical action: in the usual theories, as we know, the Schr\"odinger equation (4.6) that is derived from the classical action describing world line, is obtained by replacing the coordinates of the space-time by operators. However, we have not succeeded so far to obtain the classical action in SUSY case.

To close this section, we note that the above argument can easily be done in a superspace formalism, because each entry of the above matrix representation is nothing but the degree of superspace coordinate $\theta$. As is well known, instead of (4.3), we may use the SUSY generators in a superspace formalism in terms of chiral superfield in the following form 

\begin{align*}
	& Q_{\alpha}=\frac{\partial}{\partial \theta^{\alpha}}
	-i\sigma_{\alpha \dot{\alpha}}^{\mu}\bar{\theta}^{\dot{\alpha}}
	\partial_{\mu}, \quad \bar{Q}_{\dot{\alpha}}=	
	-\frac{\partial}{\partial \bar{\theta}^{\dot{\alpha}}}
	+i\theta^{\alpha}\sigma_{\alpha \dot{\alpha}}^{\mu}\partial_{\mu}.
\end{align*}

\noindent By making use of the Lorentz invariant inner product

\begin{align*}
	\langle \Phi |\Phi \rangle =\int d^3\vec{x} \> \Phi^{\dagger}
	\theta \sigma^0 \bar{\theta}\Phi
\end{align*}

\noindent
where we inserted $\theta\sigma^0\bar{\theta}$ instead of the matrix I in equations (4.4) and (4.5). In this way we go on the argument in the superspace formalism exactly parallel to our previous formalism where the matrix representation is used.

%%%%%%%%%%%%%%%%%%%%%%%%%%%%%%%%%%%%%%%%%%%%%%%%%%%%%%%%%%%%%%%%%%%%%%%%%%%%%%%
\section{Conclusions}

\vspace{0.5cm}

In this paper, we have proposed the idea that the boson vacuum forms a sea, like the Dirac sea for fermions, in which all the negative energy states are filled. This was done by introducing a double harmonic oscillator, which stems from an extension of the concept of the wave function. Furthermore, analogous to the Dirac sea where due to the exclusion principle each negative energy state is filled with one fermion, in the boson case we also discussed a modification of the vacuum state so that one could imagine two types of different vacuum fillings for all the momenta. The usual interpretation of an anti-particle, as a hole in the negative energy sea, turns out to be applicable not only for the case of fermions but also for that of bosons. Thus, we have proposed a way of resolving the long-standing problem in field theory that the bosons cannot be treated analogously to the Dirac sea treatment of the fermions. Our presentation relies on the introduction of the double harmonic oscillator, but that is really just to make it concrete. What is really needed is that we formally extrapolate to have negative numbers of particles, precisely what is described by our ``double harmonic oscillator", which were extended to have negative numbers of excitation quanta. Supersymmetry also plays a substantial role in the sense that it provides us with a guideline for how to develop the method. In fact, our method is physically very natural when we consider supersymmetry, which, in some sense, treats bosons and fermions on an equal footing. 

Our picture of analogy between fermion and boson sea description is summarized by Table 2.

\begin{center}
	\begin{tabular}{|c|c|c|} \hline
	& \multicolumn{2}{|c|}{\textbf{\large Fermions}} \\
	& \small{positive single particle energy} & 
	\small{negative single particle energy} \\
	& $E>0$ & $E<0$ \\ \hline
	& & \\
	empty $||\tilde{0}_+\rangle$ & true & not realized in nature \\
	& & \\ \hline
	& & \\
	filled $||\tilde{0}_-\rangle$ & not realized in nature & true \\
	& & \\ \hline
	\multicolumn{3}{c}{} \\ \hline
	& \multicolumn{2}{|c|}{\textbf{\large Bosons}} \\
	& \small{positive single particle energy} & 
	\small{negative single particle energy} \\
	& $E>0$ & $E<0$ \\ \hline
	``sector with bottom" & & \\
	analogous to ``empty" & true & not realized in nature \\
	$||0_+\rangle$ & & \\ \hline
	``sector with top" & & \\
	analogous to ``filled" & not realized in nature & true \\
	$||0_-\rangle$ & & \\ \hline
	\multicolumn{3}{c}{} \\
	\multicolumn{3}{c}{Table 2: Analogy between fermion and boson sea 
	description} \\
	\end{tabular}
\end{center}

Next, we presented an attempt to formulate the supersymmetric quantum mechanics. We found that our representation of SUSY generators in the matrix form is properly constructed only when we utilize the equations of motion (Schr\"{o}dinger equations) and can be interpreted as the SUSY generators in the second quantization. Using these generators, the 1 particle states of boson and fermion are related within the positive and negative energy sectors respectively. Therefore, we could conclude that our SUSY generators in the matrix form are useful to formulate the supersymmetric and relativistic quantum mechanics.

\vspace{0.5cm}

\noindent \underline{ Acknowledgement }

\vspace{0.25cm}

Y.H. is supported by The 21st century COE Program ``Towards a New Basic Science; Depth and Synthesis", of the Department of Physics, Graduate School of Science, Osaka University. This work is supported by Grants-in-Aid for Scientific Research on Priority Areas, Number of Areas 763, ``Dynamics of strings and Fields", from the Ministry of Education of Culture, Sports, Science and Technology, Japan.

%%%%%%%%%%%%%%%%%%%%%%%%%%%%%%%%%%%%%%%%%%%%%%%%%%%%%%%%%%%%%%%%%%%%%%%%%%%%%%%

\end{document}